\documentclass[pra,twocolumn,showpacs,preprintnumbers,amsmath,amssymb,superscriptaddress]{revtex4}

\usepackage{graphicx}
\usepackage{bm}
\usepackage{physics}

\begin{document}

\title{On the role of the measurement apparatus in quantum measurements}

\author{
Muzaffar Qadir Lone$^{1}$, Chris Nagele$^{2}$, Brad Weslake$^{2}$ and Tim Byrnes$^{3,2,4,5,6}$}

\address{
$^{1}$Department of Physics, University of Kashmir, Srinagar-190006, India\\
$^{2}$New York University Shanghai, 1555 Century Ave, Pudong, Shanghai 200122, China\\
$^{3}$State Key Laboratory of Precision Spectroscopy, School of Physical and Material Sciences,East China Normal University, Shanghai 200062, China\\
$^{4}$NYU-ECNU Institute of Physics at NYU Shanghai, 3663 Zhongshan Road North, Shanghai 200062, China\\
$^{5}$National Institute of Informatics, 2-1-2 Hitotsubashi, Chiyoda-ku, Tokyo 101-8430, Japan\\
$^{6}$Department of Physics, New York University, New York, NY 10003, USA
}




\begin{abstract}
We study the extent to which the outcomes of a quantum measurement can be manipulated by changing the state of the measurement apparatus. The measurement process is modeled as decoherence induced by the experimenter, to gain knowledge about a particular system.  The measurement apparatus is assumed to contain a large number of degrees of freedom, and the measurement outcomes are obtained by looking at the long interaction time limit.  We study two cases which show contrasting behaviour. With a fixed axis coupling, the measurement is performed along the pointer basis with a high degree of robustness, for a wide variety of bath states.  In a second model with Heisenberg interactions, the measurement outcomes can be altered considerably by changing the state of the bath.  
\end{abstract}



\maketitle

\section{Introduction}
Quantum measurement is generally regarded as an inherently random and irreversible process. In the Copenhagen interpretation of quantum mechanics, a wavefunction $ \psi(x) $ on measurement of its position $ x $, collapses to a precisely specified position, with probability density $ |\psi(x)|^2 $.  The nature (or necessity) of this collapse has always been a controversial point in quantum theory, and to this date there is no way to explicitly derive this collapse -- the so-called ``measurement problem'' \cite{auletta2000foundations, schlosshauer2005decoherence,allahverdyan2013understanding}. The peculiar way that measurements work in quantum theory in more recent years is now considered a resource that can be exploited for quantum technologies.  Quantum random number generators \cite{herrero2017quantum,ma2016quantum} work on the principle that quantum measurements are fundamentally random, and have potential applications in Monte Carlo simulations.  Quantum cryptography schemes such as the BB84 protocol are based on the randomness and repeatability of measurements for non-orthogonal states \cite{bennet1984quantum,brassard2000limitations,schiavon2016experimental}. In this way, a robust understanding of quantum measurements are not only a question for understanding the foundations of quantum mechanics, it is fundamental to the next generation of quantum technologies. 

The theory of decoherence has given a more quantitative understanding of the quantum-classical transition \cite{schlosshauer2005decoherence,zurek2003decoherence, allahverdyan2013understanding}.  The pioneering ideas of Zurek and co-workers have pointed out the generic phenomenology of when a single quantum system interacts with many others.  In this picture, one can then understand measurements as being ``man-made'' decoherence.  When a measurement is being performed, an initially pure state of the system becomes entangled with the apparatus (i.e. the reservoir), under standard Schrodinger evolution of the states.  On tracing out the apparatus one then obtains a mixed state with probabilities that coincide with the Born rule \cite{weinberg2016what, distler2017von}.  The decoherence view of quantum measurements thus gives a way of explicitly calculating the inner workings of a quantum measurement. This has made it an overwhelmingly popular approach particularly in the field of quantum information, where direct experiments have been observed predicting its effects \cite{brune1996observing}. In addition, decoherence has been proposed as a useful experimental tool, including a method of direct detection of dark matter, whereby dark matter particles interact with atoms in quantum superpositions and destroy the phase information in a detectable interaction \cite{riedel2013direct}.

The decoherence approach to quantum measurement does not solve the measurement problem as it does not explicitly predict a random collapse to one of the states.  This can be seen immediately from the fact that the Schrodinger equation -- which is the basis of decoherence theory -- is a deterministic equation. This then raises the question of in what sense quantum measurements are genuinely random.  To give a classical analogy, the outcome of a ball on a roulette wheel is predictable if the dynamics of the ball could be carefully tracked and modeled, owing to the fact that classical physics is completely deterministic.  In this same way, we might argue that if we take the decoherence view of quantum measurement, that perhaps at least part of the randomness originates from incomplete knowledge of an intractably large number of degrees of freedom of the measurement apparatus.  Is there a possibility that the deterministic evolution Schrodinger equation would give a similar conclusion as the classical example, and part of the randomness of  quantum measurements originate from the randomness of the state of the mesurement apparatus? It is important here to point out that we are not suggesting anything to do with hidden variable theory; rather, we say that the measurement result can depend both on the state of the system and the state of the apparatus, and at least some part of the randomness originates from the lack of knowledge of the apparatus.

\begin{figure}[t]
\centering\includegraphics[width=\columnwidth]{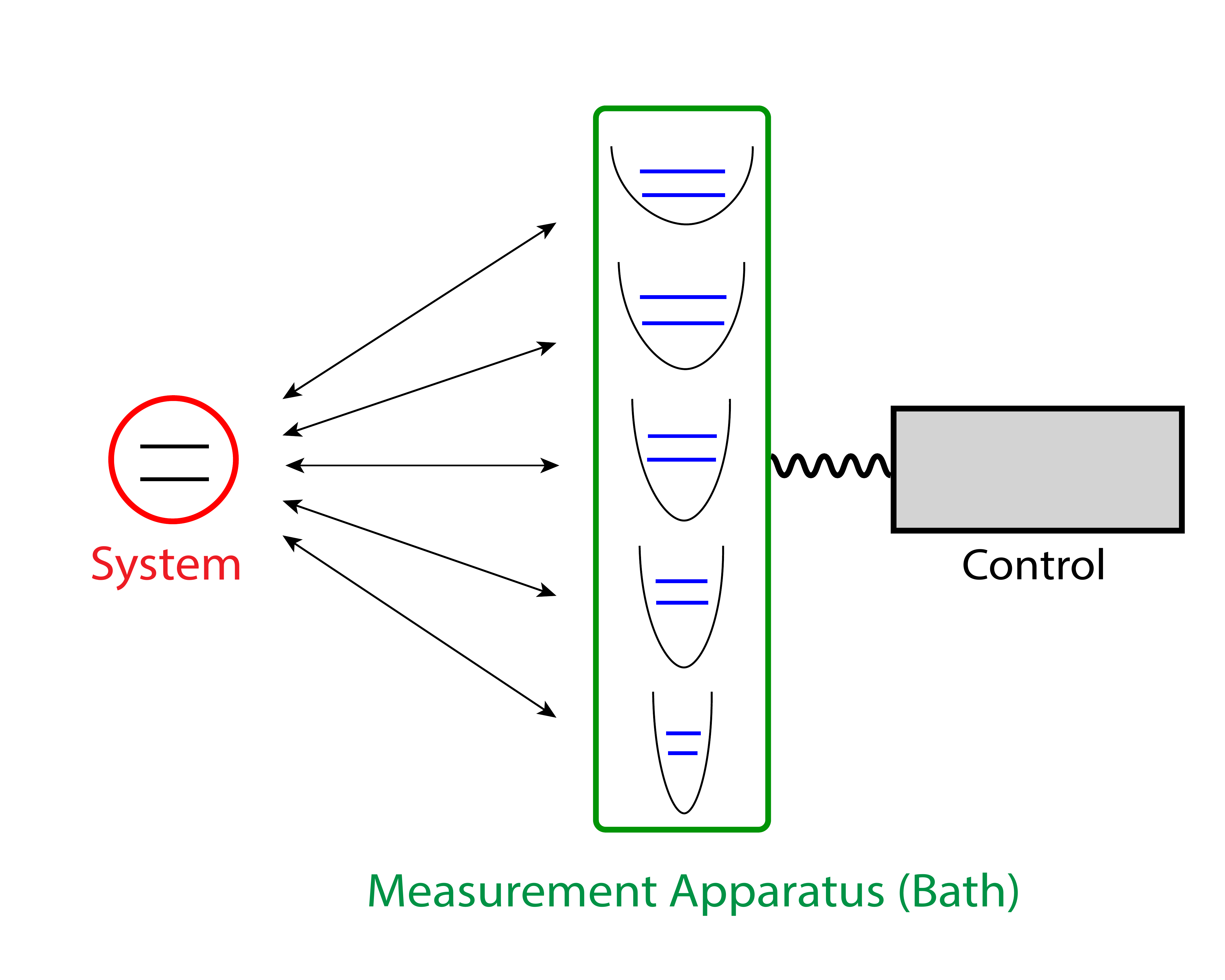}
\caption{The scenario considered in this paper. A system to be measured is coupled to the measurement apparatus with a fixed coupling $ H_I $.  The measurement apparatus consists of many degrees of freedom of different energies, and acts as a bath.  The state of the bath is assumed to be controllable by some external means.   }
\label{fig1}
\end{figure}

To answer the above question, we first need to establish what degree the state of the measurement apparatus can affect the outcome of a measurement. Specifically, in this paper we examine the configuration as shown in 
Fig. \ref{fig1}.  The system (which for simplicity we consider to be a qubit) comes in contact with the measurement apparatus, consisting of a large number of degrees of freedom.  The system is then evolved for a sufficiently long time such that steady-state is achieved. Then to what extent can the subsequent dynamics of the system be manipulated by just changing the state of the measurement apparatus? We point out that this is a weaker question than the one posed in the previous paragraph, since this can be answered by simply looking at the density matrix of the system.  A full answer of the question in the previous paragraph would require a satisfactory solution to the measurement problem -- something which we do not attempt to resolve in this paper.  Nevertheless, our results show that a considerable degree of control can be performed under these assumptions. 

To illustrate the degree of control that can be attained, we investigate and compare two models of measurement.  In both cases we model measurement as a system coupled to a bath containing many degrees of freedom.  In the first model, we take a bosonic bath coupled with a $ \sigma^z $ interaction.  We consider the case where the state of the bath is a pure coherent state, such that there is full knowledge of the bath. This type of Hamiltonian has been investigated in numerous works \cite{schlosshauer2005decoherence,zurek2003decoherence,ilookeke2014}, and several works have investigated the time scale of decoherence \cite{brasil2013how,brasil2014efficient}. In the second model, we consider a Heisenberg coupling and take the bath degrees of freedom to be a spin bath.   This model, called the ``central spin model'' has been investigated more in the context of understanding decoherence of a quantum dot interacting with a bath of nuclear spins \cite{prokofev2000theory,khaetskii2002electron, coish2004hyperfine, erlingsson2004evolution,coish2010free,ishida2013photoluminescence,wu2016decoherence,bhattacharya2017exact}. While this is a somewhat unconventional model of measurement, the Heisenberg interaction is a naturally occurring type of interaction that could be potentially used to couple the measurement apparatus to the system.  Recently, related questions regarding the Markiovianity of a quantum measurement was investigated in Ref. \cite{glick2017markovian}, although 
a different approach to the decoherence model is taken.

This paper is structured as follows. We consider two main models which illustrate two contrasting cases where the measurement dynamics is  insensitive to the state of the bath  (Sec. \ref{robust}), and another where the state of the bath affects the outcome strongly (Sec. \ref{variable}).  For the first model, we first examine the simplest case under a  Markovian assumption and a thermal bath (Sec. \ref{markovian}).  We then show that the conclusions still hold under a more general non-Markovian formalism and a state dependent bath state (Sec. \ref{generalbath}).  For the second model, we solve the model numerically for a small system for a fully polarized bath (Sec. \ref{fullypol}).  We then solve this model for a large number of bath spins by mapping the bath particles to hard core bosons (Sec. \ref{largespins}).  Finally, we give our conclusions in Sec. \ref{conc}.

\section{Case I: Fixed-axis measurements}

\label{robust}

\subsection{The model}

We first examine the case where the state of the measurement apparatus has little effect on the measurement outcome.  This is the limit where the measurement is very robust, except for some isolated examples which we also discuss. The model we first consider is the standard model of a qubit interacting with a bosonic reservoir with a $ \sigma^z $ coupling:
\begin{align}
H = H_S + H_B + H_I
\label{generalham}
\end{align}
where the system, bath, and interaction Hamiltonians are 
\begin{align}
H_S & = \frac{\omega_0}{2} \sigma^z \nonumber \\
H_B & = \sum_{k} \omega_k b_k^\dagger b_k  \nonumber \\
H_I & =  \sigma^z \sum_k (g_k  b_k + g_k^* b_k^\dagger) ,
\label{spinboson}
\end{align}
where  $\omega_0$ is the energy splitting of the qubit, $b_k$ is the anhilation operator of the $k$th bath mode, and 
where we use units such that $ \hbar = 1 $. We will consider this model in a variety of different contexts, both in the Markovian and non-Markovian approximations.  In the context of decoherence, the model (\ref{generalham}) and (\ref{spinboson}) is used to study dephasing of a qubit. Here, we take this to be a model of measurement in the $ \sigma^z $-basis. We are in particular interested in the effect of the choice of quantum state of the reservoir, and how the long-time dynamics are affected by this choice.

\subsection{Thermal bath and Markovian assumption}
\label{markovian}

Here we review the treatment of a measurement as a decoherence process for the model above.  The solution of (\ref{generalham}) under the Markovian assumption is a standard derivation and we do not reproduce it here.  For details, we refer the reader to Ref. \cite{schlosshauer2007decoherence} for details.  The initial state of the system and the bath is taken to be
\begin{align}
\rho(0) = | \psi  \rangle \langle \psi   | \otimes \rho_B (T)
\label{initialstate}
\end{align}
where the initial state of the system is
\begin{align}
| \psi \rangle = a |0 \rangle + b | 1 \rangle
\label{initialsystem}
\end{align}
with $ |a|^2 + |b|^2 = 1 $ and the bath is 
\begin{align}
\rho_B (T) = \frac{e^{-H_B/k_B T }}{Z}
\label{thermalbath}
\end{align}
with $ k_B $ the Boltzmann constant and $ Z $ is a normalization factor. One can then follow a standard procedure under the assumptions of a Markiovianity to obtain 
\begin{align}
\frac{d \rho_S}{dt} = \frac{\gamma(T)}{2} {\cal L} [\sigma^z,\rho_S]
\label{mastereq}
\end{align}
where the Lindblad superoperator is defined as
\begin{align}
{\cal L} [O,\rho] \equiv 2 O \rho O^\dagger - O^\dagger O \rho - \rho O^\dagger O 
\end{align}
and the decay constant $\gamma(T) $ depends on the bath temperature. A more detailed expression for $\gamma (T) $ can be given with respect to microscopic parameters of the original Hamiltonian, but this is irrelevant for the long-time dynamics.  We refer the reader to Ref. \cite{schlosshauer2007decoherence} for details of the solution.  

Under these initial conditions the master equation (\ref{mastereq}) can be straightforwardly solved to give the time dynamics
\begin{align}
\rho_S (t) = \left(
\begin{array}{cc}
| a|^2 &a b^*  e^{-\gamma(T) t} \\
a^* b e^{-\gamma(T) t} & | b |^2 
\end{array}
\right)
\end{align}
Assuming that the measurement timescale $ t_{\text{meas}} $ is much longer than decay timescale, i.e. $ \gamma(T)  t_{\text{meas}} \gg 1 $ and that $  \gamma(T) \ne 0 $, the density matrix always converges to 
\begin{align}
\rho_S (t \rightarrow \infty ) = \left(
\begin{array}{cc}
|a|^2 & 0\\
0 & |b|^2 
\end{array}
\right)  .
\label{decoheredstate}
\end{align}
Thus we obtain a statistical mixture of $ | 0\rangle $ and $ | 1 \rangle $, with probabilities $ |a |^2 $ and $ |b |^2 $ respectively, consistent with the Born rule for quantum measurement.  

The above is equivalent to projecting the states with measurement operators
\begin{align}
\rho_S \rightarrow \rho_S' = \sum_i M_i \rho_S M_i^\dagger
\end{align}
which is this case are
\begin{align}
M_0 & = \frac{1}{2} ( 1 + \sigma^z) = | 0 \rangle \langle 0 | \nonumber \\
M_1 & = \frac{1}{2} ( 1 - \sigma^z) = | 1 \rangle \langle 1 | .
\end{align}
We can understand this notion in the context of the so-called ``pointer basis'', where the quantum measurement picks out the $z$ basis as the most fit basis for the measurement \cite{zurek2003decoherence}. When the measurement is conducted in bases other than the $z$ basis, the information associated with the qubit in that basis is destroyed. Because of the $\sigma^z$ coupling, however, measurements in the $z$ basis are resilient to this affect and so the information survives the measurement. In this way, after a short period of time, the only information left is in the $z$ basis, making this the ``pointer basis'' which corresponds to the diagonal entries of $\rho_S$ above.

The fact that only the diagonal components survive can be derived from the dynamics (\ref{mastereq}) by setting the left hand side to zero, assuming steady state conditions:
\begin{align}
\sigma^z \rho_S  \sigma^z = \rho_S .
\end{align}
Expanding the density matrix as $ \rho_S = \sum_{\sigma \sigma'} \rho_S (\sigma, \sigma') |\sigma \rangle \langle \sigma'| $ for $ \sigma, \sigma =\{0,1\} $, this yields the conditions for diagonal and off-diagonal elements 
\begin{align}
\rho_S (\sigma, \sigma)  & = \rho_S (\sigma, \sigma)  \nonumber \\
\rho_S  (\sigma, \bar{\sigma}) & = - \rho_S  (\sigma, \bar{\sigma})
\end{align}
where $ \bar{\sigma} = 1 - \sigma $. The first condition is always satisfied, hence there is no restriction on the diagonal elements.  The second condition can only be satisfied if $ \rho(\sigma, \bar{\sigma})  = 0 $, hence all the off-diagonal elements must be zero at steady state given the dynamics (\ref{mastereq}).

We can already notice one thing about the role of the state of the measurement apparatus here.  In this simple model the state of the bath could only be changed with one parameter, the temperature $ T $.  Neglecting the transient dynamics, which are not usually observed in a quantum measurement, the state always approaches the same state (\ref{decoheredstate}) regardless of the choice of $ T $. This is true for any temperature excluding $ \gamma(T) = 0  $, which occurs for zero temperature.  Thus in this case we conclude that there is effectively no role of the state of the bath in terms of the measurement, since in the long-time limit any choice of bath would yield the same state.

\subsection{State dependent bath and non-Markovianity}
\label{generalbath}

Under the standard assumptions of Markovianity and a thermal bath as given in the previous section, we found that there is little influence of the bath state on the long-time behaviour of the measurement outcome.  One possibility for the lack of dependence is the assumption that was made for the initial state of the system.  In (\ref{initialstate}), the initial state is assumed to be a completely uncorrelated state.  But if the bath was prepared dependent upon the state of the system, perhaps this can be made to influence the outcome of the measurement.  Put another way, if we tailor a different bath for different states of the system, does this influence the measurement in any way?  In this section we study the example of a state dependent bath and at the same time, we also extend the result to a non-Markovian context to see if this provides any difference to the results of the previous section. 

We will assume a particular type of initial state of the bath and system as a case study to explore the above scenario.  Before $ t = 0 $, we initially start in the state for the combined system and bath that are thermal states of the {\it whole} system 
\begin{align}
\rho  = \frac{e^{-H/k_B T }}{Z}
\label{thermalbath2}
\end{align}
where $ Z $ is the partion function. 
This should be compared to (\ref{thermalbath}) where the thermal state was only with respect to the bath states.  Now we would like to prepare the state of the system in a similar state to that considered in the previous section.  To do this we assume that the state of the system is known in advance, and a projective measurement is made on the system and bath. For the system the projection operators are
\begin{align}
P_S  = \{ | \psi \rangle \langle \psi |, | \psi_\perp \rangle \langle \psi_\perp | \}
\end{align}
where $ | \psi \rangle $ is the same initial state as (\ref{initialsystem}) and  $ | \psi_\perp  \rangle = b^* | 0 \rangle - a^* | 1 \rangle $ is the state that is orthogonal to this state. The bath states are also projected onto the set of coherent states
\begin{align}
P_B (\alpha_k)= | \alpha_k \rangle \langle \alpha_k | 
\end{align}
where $|\alpha_k\rangle$ is a  coherent state of the $ k$th mode of the bath. Now we postselect on the state  $ | \psi \rangle $ such that the system is ensured to have the state 
\begin{align}
\rho_S (0) = |\psi\rangle \langle \psi|  .
\end{align}
The bath states then take a form
\[\hspace*{-3cm} \rho_B^{\psi} (0)= \frac{1}{Z}  \int \prod_k d^2 \alpha_k*\]
\begin{eqnarray}
\hspace*{1cm}
 \left( \langle \psi  | \otimes \langle \alpha_k | e^{-\beta H} |  \psi \rangle \otimes |  \alpha_k \rangle \right)
 |\alpha_k \rangle \langle \alpha_k|
\end{eqnarray}
where there is a dependence on the state of the system because the original state (\ref{thermalbath2}) were thermal states in the space of the system and bath.  The initial state of the whole system is then 
\begin{align}
 \rho(0) = \rho_S (0) \otimes  \rho_B^{\psi} (0) .
\end{align}
The primary difference to the case examined in the previous section is that the bath state according to this preparation depends in a non-trivial way on the system state.  Remarkably, despite the additional complications brought about due to this initial state, it is possible to solve this case even under non-Markovian assumptions as we show below.  

The details of the calculations are rather cumbersome hence we show the full derivation in the Appendix.  The final result is
\begin{eqnarray}
\label{DEN}
 \rho_S(t) = \begin{pmatrix}
|a |^2 &a b^* e^{-i(\omega_0 t +\chi(t))}e^{-\gamma(t)}\\
a^* b e^{i(\omega_0 t+ \chi(t))}e^{-\gamma(t)}&|b|^2
\end{pmatrix}
\end{eqnarray}
where 
\[\hspace*{-2cm} \gamma(t)=\int_0^\infty d\omega J(\omega)\frac{1-\cos\omega t}{\omega^2}\coth\frac{\beta\omega}{2}-\]
\begin{eqnarray}
\hspace*{1cm}
\label{dec}
\frac{1}{2} 
\ln \Bigg[ 1-\frac{(1-\langle\sigma_z\rangle^2)\sin^2[\Phi (t)]}{\Big(\cosh(\beta \omega_0/2)-
\langle\sigma_z\rangle \sinh(\beta \omega_0/2)\Big)^2} \Bigg] \nonumber\\
\end{eqnarray}
and
\begin{eqnarray}
Phi(t)&=&\int_0 ^\infty d\omega J(\omega)\frac{\sin(\omega t)}{\omega^2}\\
\tan \chi(t) &=& \frac{\sinh(\beta \omega_0/2) -
\langle\sigma_z\rangle \cosh(\beta \omega_0/2)}{\cosh(\beta \omega_0/2)-
\langle\sigma_z\rangle \sinh(\beta \omega_0/2)}\tan \Phi (t)  .
\end{eqnarray}
and $J(\omega)$ is the spectral density function and $ \langle\sigma_z\rangle =  |a |^2 - | b|^2 $.  The first term in (\ref{dec}) is the usual decoherence
term and is due to vacuum and thermal fluctuations of the bath. The second term is due to the  initial correlation
between bath and the system.

In the long-time limit, the decoherence factors in (\ref{DEN}) collapse the off-diagonal terms to zero.  We thus again have
\begin{eqnarray}
\label{DEN2}
 \rho_S(t \rightarrow \infty) = \begin{pmatrix}
|a |^2 & 0 \\
0&|b|^2 .
\end{pmatrix}
\end{eqnarray}
This was the same result that was obtained in the previous section under a Markovian assumption and a thermal bath. These results are suggestive that a quantum measurement with the general Hamiltonian (\ref{generalham}) always gives a state of the  form (\ref{DEN2}) in the long time limit, regardless of the choice of bath state.  One can view this as guaranteed from the symmetry of the Hamiltonian (\ref{spinboson}).  Since the whole Hamiltonian commutes with the $ \sigma^z $ operator, the diagonal states of the density operator $ | 0 \rangle \langle 0 | $ and $ | 1 \rangle \langle 1 | $ are an invariant of the dynamics and are guaranteed to remain the same.  The only thing that can happen is then a modification of the off-diagonal components, which in the context of a bath are in most likely to decay for long timescales. We thus conclude for this type of interaction that the measurement scheme is extremely robust in the sense that apart from some extreme states that can cause different dynamics, virtually any state of the bath will cause a decay to the (\ref{decoheredstate}). 

The above is the most likely situation for a bath states which are not particularly carefully chosen.  Are there other scenarios that could occur for more carefully chosen bath states? The answer to this is certainly yes. For example, if the bath states are chosen to be an eigenstate of the Hamiltonian (e.g. an infinitely squeezed state of the appropriate phase $ (e^{i \phi} b_k + e^{-i \phi} b_k^\dagger) | \phi \rangle \propto | \phi \rangle $, then the system and the bath never become entangled, and the effect of the interaction Hamiltonian is to simply apply a rotational phase $ H_I \propto \sigma^z $.  Then if the bath states are instantaneously changed to a completely mixed state, then again the system and the bath never couple, and the effect of the bath is zero.  In this way the ``measurement apparatus'' can be converted to a (admittedly highly contrived) quantum control scheme, and no measurement is performed. No randomness ever results as the state of the system remains pure.   Of course this is a highly atypical situation, but we note that this is still allowable according to the rules of our game, where the state of the measurement apparatus is changed.  This type of situation would be the exception to the general conclusion as given above, that most states of the bath collapse to (\ref{decoheredstate}).

\section{Case II: Variable measurements}
\label{variable}

\subsection{The model}

The results of the previous section illustrate the conventional view of quantum measurements, that the state of the measurement apparatus does not play a significant role to the measurement outcomes.  This is fortunate, as after all an experimenter typically does not carefully manipulate the quantum state of the apparatus. Hence there exists an implicit assumption that the measurement will give the same results regardless of the broad conditions that the detector is placed in. 

In this section, we give a counterexample to the previous scenario, where the state of the measurement apparatus has a very significant effect on the measurement outcome.  Our model is the central spin model which models a quantum dot interacting with a bath of nuclear spins \cite{prokofev2000theory,khaetskii2002electron, coish2004hyperfine, erlingsson2004evolution , coish2010free,wu2016decoherence,bhattacharya2017exact}.   In the central spin model an electron spin interacts via a Heisenberg interaction to its surrounding nuclear spins and is given by the Hamiltonian
\begin{align}
H = H_S + H_B + H_I
\end{align}
where
\begin{align}
H_S & = \frac{\omega_0}{2}  \sigma_0^z \nonumber \\
H_B & = \frac{1}{2} \sum_{k=1}^N \omega_k \sigma^z_k \nonumber \\
H_I & = \frac{  \bm{\sigma}_0}{2} \cdot \sum_{k=1}^N g_k \bm{\sigma}_k .
\label{hamspinbath}
\end{align}
Here we have labeled the spins by $ \bm{\sigma}_k = (\sigma^x_k, \sigma^y_k,\sigma^z_k) $, where the system spin corresponds to $ k = 0 $ and the bath spins are $ k \in \{1, \dots, N \} $. The magnetic field on the system spin creates an energy difference of $ \hbar \omega_0 $, and the bath spins are assumed to have a distribution of frequencies $  \omega_k $. The central spin model has been observed to have a strong dependence on the state of the bath, although to our knowledge it has not been considered in the context of quantum measurements before \cite{schliemann2002spin,coish2004hyperfine}.

\subsection{Fully polarized bath}
\label{fullypol}

We now examine the case where the state of the measurement apparatus is fully polarized.  For our purposes the variation of the coupling $ g_k $ is not necessary to observe the effect of the bath state, hence in this section we will set this to a constant $ g_k = g $.  
The initial state of the whole system that we consider is a pure state of the form
\begin{align}
| \psi (0) \rangle = ( a | 0 \rangle_S + b | 1 \rangle_S ) \otimes
\prod_{k=1}^N \left( c | 0 \rangle_{Bk} + d | 1 \rangle_{Bk} \right) ,
\label{initialstatespin0}
\end{align}
where $ a,b,c,d $ are arbitrary coefficients such that $ |a|^2 + |b|^2 =1 ,  |c|^2 + |d|^2 =1  $, and the  state of the $ k $th bath spin is labeled by the subscript $ Bk $. Now we point out that the way that we have written (\ref{hamspinbath}) and (\ref{initialstatespin0}) there are two preferred axis directions. In the Hamiltonian there is the axis specifying the energy, given by the presence of the $ \sigma^z $ operators.  In the state (\ref{initialstatespin0}) there is the direction of the polarization of the bath.  In principle these can be different,  but here we will assume that the bath spins are aligned with the energy direction (which might explain why they are polarized in the first place).  Making a rotation 
\begin{align}
| 0 \rangle & = c^* | 1 \rangle' + d | 0\rangle' \nonumber \\
| 1 \rangle & = d^* | 1 \rangle' - c | 0\rangle' ,
\label{basistrans}
\end{align}
the state in the new basis is
\begin{align}
| \psi (0) \rangle' = ( \alpha | 1 \rangle_S' + \beta | 0 \rangle_S' ) \otimes \prod_{k=1}^N  
| 1 \rangle_{Bk}' ,
\label{initialstatespin}
\end{align}
where 
\begin{align}
\alpha & = ac^* + b d^* \nonumber \\
\beta & = a d - b c .
\end{align}
We thus take the $ | 0 \rangle', | 1 \rangle' $ to be the eigenstates of the $ \sigma^z $ operator. 

The key point to be noticed here is that in performing the basis transformation (\ref{basistrans}), the interaction Hamiltonian $ H_I $ is always left invariant because the Heisenberg interaction $  \bm{\sigma}_0 \cdot \bm{\sigma}_k $ is invariant under a basis transformation.  This is a crucial difference to the model examined in Sec. \ref{robust} where the form of the interaction would also change under the transformation.  This allows for the possibility that the state of the bath can affect the dynamics of the measurement.

\subsection{Numerical solution for small number of spins}
\label{numerical}

Henceforth dropping the dashes in the states  $ | 0 \rangle' \rightarrow | 0 \rangle, | 1 \rangle' \rightarrow | 1 \rangle $, we may solve the time dynamics of the Hamiltonian (\ref{hamspinbath}) with the initial conditions (\ref{initialstatespin}) numerically in a straightforward way.  For the initial condition (\ref{initialstatespin}), and using the fact that the Hamiltonian preserves $ \sum_{k=0}^N \sigma_k^z $, one finds that the Hilbert space has a dimension of $ N + 1 $, which can be numerically diagonalized easily \cite{schliemann2002spin}.  This is what allows for an exact solution in the central spin model \cite{khaetskii2002electron}. Firstly, the state $ | 1 \rangle_S  \otimes \prod_{k=1}^N | 1 \rangle_{Bk} $ is an eigenstate of the Hamiltonian with energy
\begin{align}
\frac{gN}{2} - \frac{1}{2} \sum_{k=0}^N \omega_k     .
\end{align}
Hence this term in the superposition (\ref{initialstatespin}) is invariant except for the accumulation of a phase.  For the other term in the superposition with a single excitation, the Hamiltonian takes a form
\[\hspace*{-1cm} H = \left(
\begin{array}{ccccc}
\omega_0 - gN & g & g & \dots & g \\
g &  \omega_1-g  & 0 & \dots & 0 \\
g & 0 & \omega_2 - g  & \dots & 0 \\
\vdots & \vdots &                    &      &   \\
g & 0 &  \dots & 0 & \omega_N  -  g 
\end{array} 
\right) \]
\begin{align}
\hspace*{3cm}
- ( \omega_0 - gN + \sum_{k=1}^N \omega_k ) \frac{I}{2}
\label{oneexcitationsector}
\end{align}
where the basis vectors are $ \sigma^-_k |1\rangle_S \otimes \prod_{k=1}^N  
| 1 \rangle_{Bk} $ in the order $ k = \{ 0,1,\dots, N \}.  $

\begin{figure}[t]
\centering\includegraphics[width=\columnwidth]{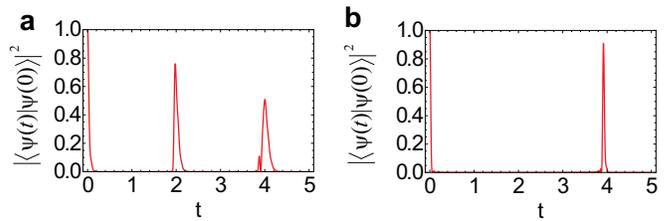}
\caption{Time evolution for a polarized bath for the central spin problem Hamiltonian (\ref{oneexcitationsector}) for (a) $ N = 50 $; and (b) $ N = 100 $.   The state of the system and bath is initially $ |\psi(0) \rangle =  | 0 \rangle \otimes \prod_{k=1}^N  
| 1 \rangle_{Bk} $, and the occupation probability of measuring in this state is calculated.  Parameters used are $ g = 4 $, $ \omega_0 = g (N-1) $, $ \omega_k/2 = 39 - 80 k/(N-1) $.  }
\label{fig2}
\end{figure}

Numerical results for typical parameters are shown in Fig. \ref{fig2}.  For the $ | 0 \rangle $ state of the system, this can be seen to decay into the bath, with most of the time the state relaxing to $ | 1 \rangle $, with negligible probability in the original $ | 0 \rangle $.  At certain times revivals are observed, where the excitation returns again to the system.  This is due to finite size effects of the simulation, for larger number of system qubits, the revival times are extended, as can be seen by comparing Fig. \ref{fig2}(a) and (b).  We can easily extrapolate that for very large system sizes $ N \rightarrow \infty $ no revivals would be present and the amplitude of the $ | 0 \rangle $  state decays irreversibly to zero.  

This can again be viewed as a measurement, since it distinguishes the states $ | 0 \rangle $ and $ | 1 \rangle $ of the system by entangling the system and the bath.  Specifically, the measurement operators corresponding to this are
\begin{align}
M_0 & = (I-\sigma^z)/2 \nonumber \\
M_1 & = \sigma^+
\end{align}
since the first case where the bath remains in $ \prod_{k=1}^N  
| 1 \rangle_{Bk} $ only occurs when the system spin is $ | 1 \rangle $ and similarly the decay of the single excitation case only occurs when the system is in the $ | 0 \rangle $ state.  The measurement operators satisfy the completeness equation $ M_0^\dagger M_0 + M_1^\dagger M_1  = I $ and thereby satisfy a valid measurement. We note that this is similar to the way a photon is detected: if no photon is present then the vacuum remains the vacuum, but if a photon is present, it is absorbed by the detector and the final state is again the vacuum.

From the discussion surrounding (\ref{initialstatespin}) it is obvious that changing the state of the bath (i.e. different parameters $ c,d $) will correspondingly cause the measurement to be made in a different basis. The $ | 0 \rangle' $ and $ | 1 \rangle' $ states were defined to be the axis of the polarization of the bath spins, according to (\ref{basistrans}).  This in turn affects the system spins, which are then written in this new basis (\ref{initialstatespin}).  The argument then follows the same logic, so that the $ | 1 \rangle' $ state in this new basis is unaffected by the bath, and the $ | 0 \rangle' $ state relaxes to $ | 1 \rangle' $.  In this way, the way the spins are polarized directly affects which axis the system is measured. 

From a physical point of view of spins interacting via the Heisenberg interaction this is obvious, as we can view the system as under the influence of a large effective magnetic field in the direction of the polarization.  The system spin then decoheres in this unique basis, due to the presence of many different modes that are present.

We note out that in the context of quantum measurements, Refs. \cite{barnea2017macroscopic,wei2008statistics} also consider Heisenberg-type couplings of the system to the apparatus.  These however give qualitatively different results to our model due to the nature of the type of baths involved.  In these works each of the $ x, y, z $ components of the interaction Hamiltonian $ H_I $ couple to a {\it different} bath mode, while our model as a Heisenberg coupling to the {\it same} bath degree of freedom.  This makes a crucial difference to the master equation, which for these works yield a master equation of the form
\begin{align}
\frac{\partial \rho_S}{\partial t}=\frac{\gamma}{2}\left( {\cal L} [\sigma^x,\rho_S]+{\cal L} [\sigma^y,\rho_S]+{\cal L} [\sigma^z,\rho_S] \right) 
\label{mewei2008}
\end{align}
which in the long time limit decoheres $ \rho \rightarrow I/2 $, where $I $ is the identity.  In this situation, we have similar results to that of Sec. \ref{robust} where the fixed point is independent of the state of the bath.

\subsection{Large number of spins and non-Markovianity}
\label{largespins}

In the previous section we numerically solved the central spin problem for a small number of spins.  To be properly considered a quantum apparatus, more realistically there should have a large number of spins. Furthermore, the couplings $ g_k $ would more generally depend on the bath spin, rather than the constant that we assumed in the previous section.  The small number of bath spins were responsible for the collapse and revival dynamics as was observed in Fig. \ref{fig2}, which normally would not be observed in a typical measurement.  In this section we derive the non-Markovian master equation and examine the long-time behaviour for various states of the bath.  

To derive the master equation we transform the bath spins in the Hamiltonian (\ref{hamspinbath}) to a system of hard core bosons, and 
evaluate the evolution starting from the initial state (\ref{initialstatespin}).  We again relegate the details of of the calculation to the Appendix. We obtain a final master equation of the form:
\begin{eqnarray}
\label{SME}
 \frac{d\rho_S }{dt} &=& -i [\tilde{H}(t),  \rho_S]
 + \Gamma_{d}(t) {\cal L } [ \sigma^z, \rho_S]
+ \Gamma_0(t) {\cal L } [ \sigma^+, \rho_S] 
 \end{eqnarray}
where
\[\hspace*{-3cm} \tilde{H}(t) = \sigma^z \left(\frac{ \omega_0 - \sum_k g_k}{2} \right) +\]

\begin{align}
\hspace*{2.5cm} \sigma^-\sigma^+  \sum_k \frac{g_k^2 (1-\cos(\omega_0-\omega_k)t)}{\omega_0-\omega_k}
\end{align}
\begin{eqnarray}
\Gamma_d(t) &=& (\sum_k g_k)^2 t \\
\Gamma_0(t) &=&  \sum_k  \frac{g_k^2 \sin(\omega_0-\omega_k)t}{\omega_0-\omega_k} . 
\end{eqnarray}
We may explicitly solve the above master equation from the initial condition (\ref{initialstatespin}) to obtain the dynamics
\begin{eqnarray}
 \rho_S(t) = \begin{pmatrix}
 |\beta|^2 G_1(t) & \alpha^* \beta G_2(t)  \\
 \alpha \beta^* G_2^*(t)  & 1-|\beta|^2 G_1(t) ,
\end{pmatrix}
\end{eqnarray}
where 
\begin{align}
G_1(t) & =e^{-\gamma_1 (t)  } \nonumber \\
G_2(t) & = e^{-2 i \gamma_d(t)} e^{-(\sum_k g_k)^2 t^2 /2} \nonumber \\
\gamma_1(t) & = 2 \sum_k g_k^2 [\frac{1-\cos(\omega_0-\omega_k)t}{(\omega_0-\omega_k)^2}) ] \nonumber \\
\gamma_d(t) & = \sum_k g_k^2[\frac{(\omega_0-\omega_k)t - \sin(\omega_0 -\omega_k)t}{(\omega_0 -\omega_k)^2}] .
\end{align}
In the long-time limit the density matrix approaches the state $ | 1 \rangle $, in agreement to the numerical evolution that was found in the previous section.  This is true regardless of the choice of the couplings $ g_k $, which were chosen to be a constant in the previous section.  The non-exponential decay is characteristic of the non-Markovian dynamics that is taken into account in our analysis.  We note that the basis that was chosen for the calculation was the rotated basis (\ref{basistrans}), where the states are defined relative to the polarization of the bath spins.  Therefore, by rotating the polarization of the bath spins, the decay would occur in a different basis $ | 0 \rangle' , | 1 \rangle' $ with the same arguments as made in the previous section.

We thus conclude again that the basis of the measurement can be selected according to the polarization of the bath spins.  The key ingredient here was the Heisenberg interaction between the system and bath, which are rotationally symmetric.  By breaking the rotational symmetry with the polarization, this can be used to change the outcome of the quantum measurement.  The conclusion that we reach in this section is therefore rather different to that of Sec. \ref{robust}. By engineering the state of the bath, it is possible to influence the outcome of a measurement. Of course, the stochastic nature of the measurement is not removed, and Born probabilities are obtained regardless of the basis choice. Nevertheless it is interesting that without any physical change in the physical characteristics of the measurement apparatus (i.e. the Hamiltonian), the outcome of a measurement can be influenced.  We note that this is similar in concept to quantum control methods involving many-body systems \cite{burgarth2010scalable,heule2010local,maruyama2017gateway,orieux2015experimental,yan2013measurement,wang2014feed}. However, for the purposes of this paper we are interested in typical quantum measurement scenarios where such exotic manipulations are not performed, hence we are content to observe that a large scope of measurements can be performed simply by changing the state of the bath, and keeping the coupling fixed.  

By engineering exotic states of the bath, can we have other dynamics as we saw for the results of Sec. \ref{robust}  A particular example that would give non-decaying dynamics is when the state of the bath is completely parallel to the system spin.   As we have already seen in Sec. \ref{numerical}, the Hamiltonian has no effect on the state.  One could then for example adiabatically change the state of the bath such as to modify the system spin. Specifically, by slowly changing the polarization, the system spin should follow the dynamics.  We again see that the ``measurement apparatus'' can perform arbitrary quantum control on the system, without entangling the bath.  This would be a deterministic evolution without introducing any randomness to the system, since the bath and system remains decoupled at all times.

\section{Summary and Conclusions}
\label{conc}

In summary, we have examined two case studies of quantum measurements displaying different behaviour.  For a measurement with a specified axis (in our case $ \sigma^z $), we found that for typical bath states all states collapse to the $ z $ basis, recovering the Born rule of quantum measurements.  Virtually any state can be chosen and similar results of the measurement would be obtained in the large bath and long time limit.  In this sense the sensitivity of the measurement to the bath is extremely low.  In contrast, for the measurements with Heisenberg interactions, the polarization direction of the bath could be used to change the basis of the measurements freely. This was due to the rotational symmetry of the Heisenberg interactions which did not specify the pointer basis. 

For both the fixed-axis and Heisenberg coupling cases, it was clear that exotic states of the bath could be potentially used to deterministically  manipulate the system.  Preparing the bath state in an eigenstate of the Hamiltonian, or by using adiabatic evolution one can realize a type of contrived quantum control using the measurement apparatus, where no randomness is introduced to the system.  Analogies exist to the roulette wheel that was discussed in the introduction.  For most configurations of the roulette wheel, the control of the ball is rather poor and essentially random behaviour results.  However, by carefully controlling the roulette wheel, there can be situations where arbitrary control of the ball can be achieved. In this way the randomness of the quantum measurement can indeed be changed according to the preparation of the state.  However, unlike the classical case, there is an inescapable randomness that can never be removed arising from the quantum correlations (e.g. entanglement) of the system and bath -- which arises in the more typical situations.  Once the system and bath are entangled, tracing out the bath always results in a stochasticity that does not arise from incomplete knowledge of the bath.  Ultimately tracking down the source of this randomness requires a resolution of the quantum measurement problem, which unfortunately remains elusive. 

Our results may be of practical significance in cases where it is crucial to obtain quantum measurements with very high precision.  For example, if the state $ (|0 \rangle + | 1 \rangle )/\sqrt{2} $ was measured using the method with the free axis as in Sec. \ref{variable}, then one would have to ensure that the state of the bath is precisely aligned in the $ | 0 \rangle $ direction.  In this sense the fixed-axis measurement of Sec. \ref{robust} would be more robust, assuming that the measurement axis was determined accurately.  In this paper we only investigated relatively simple configurations of the bath involving full or partial polarization.  A higher degree of control would be most likely possible with more sophisticated controls of the measurement apparatus.

\vskip6pt

\enlargethispage{20pt}






\onecolumngrid
\appendix

\section*{Appendix}

\subsection{Explicit calculation of (\ref{DEN})}

Suppose at time $t=0$ the state of the composite system is described by the initial density matrix $\rho(0)$,
then at time $t$ the density matrix in interaction picture is given by
\begin{eqnarray}
 \rho(t) = U(t)\rho(0)U(t)^{\dagger}
\end{eqnarray}
where $U(t)=T~e^{-i\int_0^t dt^{\prime} H_I(t^{\prime})}$ is the time evolution operator and $H_I(t)$ is the interaction
Hamiltonian. 
Our main interest is to calculate the reduced density matrix of the system by tracing out degrees of freedom of the bath:
\begin{eqnarray}
 \rho_S(t)= {\rm Tr}_B[  U(t)\rho(0)U(t)^{\dagger} ]
\end{eqnarray}
We write 
$U(t) = e^{i\phi(t)}e^{\sigma^z \hat{\Lambda}(t)}$ where $\phi(t)$
is a function of time only and $\hat{\Lambda}(t) = \sum_k[ g_k(t)b_k - g_k^*(t) b_k^{\dagger} ]$ with
$g_k(t) = g_k \frac{e^{-i\omega_k t}-1}{\omega_k}$. Therefore, we can write
\begin{eqnarray}
 \rho_S(t) &=& {\rm Tr}_B[U(t) \rho(0) U(t)^{\dagger}]\\
 &=& {\rm Tr}_B[ e^{\sigma^z \hat{\Lambda}(t)  } |\psi\rangle \langle \psi| \otimes 
 \rho_B^{\psi}  e^{-\sigma^z \hat{\Lambda}(t)}   ]
\end{eqnarray}
Here we assume the general state of the qubit as $|\psi\rangle= a |0\rangle + b |1\rangle$, we write
\begin{eqnarray}
\label{rho1}
 \rho_S(t) &=& |a|^2 |0\rangle \langle 0| + |b|^2 |1\rangle\langle 1| \\
 && + ab^{*} |0\rangle \langle1|  \frac{1}{Z} \int \prod_k d^2\alpha_k f(\alpha_k, \alpha_k^{*}, \psi) 
 {\rm Tr}_B [e^{\hat{\Lambda}(t)}  |\alpha_k \rangle \langle \alpha_k| e^{\hat{\Lambda}(t) }] +H.C
\end{eqnarray}
where $f(\alpha_k, \alpha_k^{*},\psi)= \langle \psi, \alpha_k | e^{-\beta H} |  \psi, \alpha_k \rangle $.
Next we write 
\begin{eqnarray}
 &&\!\!\!\!\!\!\!\!\!\!\!\!\!\!\!\!\!\!\!\!\!\!\!\!
 \int \prod_k d^2\alpha_k f(\alpha_k, \alpha_k^{*}, \psi) 
 {\rm Tr}_B [e^{\hat{\Lambda}(t)}  |\alpha_k \rangle \langle \alpha_k| e^{\hat{\Lambda}(t) }] \nonumber \\
 &=& \int \prod_k d^2 \alpha_k f(\alpha_k, \alpha_k^{*}, \psi) 
  \langle \alpha_k| e^{\hat{\Lambda}(t)}  e^{\hat{\Lambda}(t)}  |\alpha_k \rangle \nonumber \\ 
 &=&\  \int \prod_k d^2 \alpha_k f(\alpha_k, \alpha_k^{*}, \psi) 
  \langle \alpha_k | e^{\sum_k [ g_k(t)b-g_k^{*}(t) b_k ^{\dagger} ]}e^{\sum_k [g_k(t)b_k-g_k^{*}(t) b_k^{\dagger}] }|\alpha_k\rangle \nonumber\\
  &=&   \int\prod_k d^2 \alpha_k f(\alpha_k, \alpha_k^{*}, \psi) 
  e^{ \sum_k [-2|g_k(t)|^2 -2g_k^{*}(t)\alpha_k^{*} + 2g_k(t)\alpha_k]} 
  \label{QW}
\end{eqnarray}
where $b_k|\alpha_k\rangle= \alpha_k| \alpha_k\rangle$. Next we calculate $f(\alpha_k, \alpha_k^{*},\psi)$. 
We observe that $e^{-\beta H} |0\rangle =e^{-\frac{\beta \omega_0}{2}} e^{-\beta H_{+}}|0\rangle$ and 
$ e^{-\beta H} |1\rangle = e^{\frac{\beta \omega_0}{2}}e^{-\beta H_{-}} |1\rangle$
where $H_{\pm}= \sum_k [\omega_k b_k^{\dagger}b_k \pm (g_kb_k+g^*_kb_k^{\dagger} )]$. We write 
\begin{eqnarray}
\label{trace}
 f(\alpha, \alpha^{*},\psi) &=& \langle \psi, \alpha| e^{-\beta H} |\psi, \alpha\rangle \nonumber \\
 &=& |a|^2 \langle \alpha| e^{-\beta H_{+}} |\alpha \rangle e^{-\frac{\beta \omega_0}{2}} +
 |b|^2 \langle \alpha| e^{-\beta H_{-}} |\alpha \rangle e^{\frac{\beta \omega_0}{2}}.
 \end{eqnarray}

Next in order to evaluate the above expectation values, we define an operator 
\begin{eqnarray}
 O_{\pm}= e^{\pm\sum_k (\frac{g_k}{\omega_k}b_k^{\dagger}-\frac{g^*_k}{\omega_k} b_k)}
\end{eqnarray}
so that we have $O_{\pm} H_{\pm} O_{\pm}^{\dagger}= \sum_k [\omega_k b_k^{\dagger}b_k - 
\frac{|g_k|^2}{\omega_k}] $. Therefore we write
\begin{eqnarray}
 \langle \alpha_k| e^{-\beta H_{+}} |\alpha_k \rangle 
 &=& \langle \alpha_k| O_{+}^{\dagger} O_{+} e^{-\beta H_+}  O_{+}^{\dagger}  O_{+} |\alpha_k\rangle \nonumber\\
 &=& e^{\sum_k \frac{\beta |g_k|^2}{\omega_k}} \langle \alpha_k| O_{+}^{\dagger}  
 e^{-\beta \sum_k \omega_k b_k^{\dagger} b_k}   O_{+} |\alpha_k \rangle \nonumber\\
 &=& \prod_k e^{\frac{\beta |g_k|^2}{\omega_k}} \langle 
 \alpha_k | e^{-\frac{g_k b_k^{\dagger}}{\omega_k}} e^{\frac{g^*_k b_k}{\omega_k}} 
 e^{-\frac{|g_k|^2}{2\omega_k}} e^{-\beta \omega_k b_k^{\dagger} b_k}
 e^{\frac{g_k b_k^{\dagger}}{\omega_k}}e^{-\frac{g^*_k b_k}{\omega_k}}  
 e^{-\frac{|g_k|^2}{2\omega_k}} |\alpha_k \rangle \nonumber \\
 &=& \prod_k e^{\frac{\beta |g_k|^2}{\omega_k}-\frac{|g_k|^2}{\omega_k} 
 -\frac{g_k}{\omega_k}\alpha_k^{*} - \frac{g_k^*}{\omega_k}\alpha_k} 
 \langle \alpha_k |e^{\frac{g^*_k b_k}{\omega_k}}e^{-\beta \omega_k b_k^{\dagger} b_k}
 e^{\frac{g_k b_k^{\dagger}}{\omega_k}} |\alpha_k \rangle 
 \label{A}
\end{eqnarray}

Now we calculate $\langle \alpha_k |e^{\frac{g^*_k b_k}{\omega_k}}e^{-\beta \omega_k b_k^{\dagger} b_k}
e^{\frac{g_k b_k^{\dagger}}{\omega_k}} |\alpha_k \rangle $:
\begin{eqnarray}
 \langle \alpha_k |e^{\frac{g_k^* b_k}{\omega_k}}e^{-\beta \omega_k b_k^{\dagger} b_k} 
 e^{\frac{g_k b_k^{\dagger}}{\omega_k}} |\alpha_k \rangle &=&
 \int d^2\gamma~\int d^2 \lambda ~\langle \alpha_k | e^{\frac{g^*_k }{\omega_k}b_k} | \gamma \rangle
 \langle \gamma | e^{- \omega b_k^{\dagger} b_k} |\lambda \rangle \langle \lambda | 
 e^{\frac{g_k}{\omega}b_k^{\dagger}}|\alpha_k\rangle \nonumber \\
 &=& \int d^2\gamma~\int d^2 \lambda~ e^{[\frac{g^*_k}{\omega_k} \gamma 
 + \gamma^{*} \lambda e^{-\beta \omega_k} + \frac{g_k}{\omega_k} \lambda^{*} + \alpha_k^{*} \gamma + \lambda^{*} \alpha_k]}\\
 &=& \exp[|\alpha_k+ \frac{g_k}{\omega_k}|^2 e^{-\beta \omega_k}]
\end{eqnarray}
where we have used $\langle \alpha_k| \gamma\rangle = e^{\alpha_k^{*} \gamma}$ and $\langle \gamma| e^{-\beta \omega_k b_k^{\dagger}b_k} |\lambda\rangle =
e^{\gamma^{*} \lambda e^{-\beta \omega_k }}$.

Using this result in equation (\ref{A}), we get 
\begin{eqnarray}
\label{H+}
 \langle \alpha_k| e^{-\beta H_{+}} |\alpha_k \rangle =
 \prod_k e^{\frac{\beta |g_k|^2}{\omega_k}}  e^{-\frac{|g_k|^2}{\omega_k^2}[1-e^{-\beta \omega_k}]} e^{|\alpha_k|^2 e^{-\beta \omega_k } -\frac{g^*_k \alpha_k+ g_k \alpha_k^*}{\omega_k}(1-e^{-\beta \omega_k})  }
\end{eqnarray}
Similarly we can calculate $\langle \alpha_k | e^{-\beta H_{-}} |\alpha_k \rangle$
\begin{eqnarray}
\label{H-}
 \langle \alpha_k| e^{-\beta H_{-}} |\alpha_k \rangle =
 \prod_k e^{\frac{\beta |g_k|^2}{\omega_k}}  e^{-\frac{|g_k|^2}{\omega_k^2}[1-e^{-\beta \omega_k}]} e^{|\alpha_k|^2 e^{-\beta \omega_k } +\frac{g^*_k \alpha_k+ g_k \alpha_k^*}{\omega_k}(1-e^{-\beta \omega_k})  }.
\end{eqnarray}
Using equations (\ref{H+}), (\ref{H-}), (\ref{trace}), we can write equation (\ref{QW})
\begin{eqnarray}
\label{fr}
 &&\!\!\!\!\!\!\!\!\!\!\!\!\!\!\!\!
 \int \prod_k d^2\alpha_k f(\alpha_k, \alpha_k^{*}, \psi) 
 {\rm Tr}_B [e^{\hat{\Lambda}(t)}  |\alpha_k \rangle \langle \alpha_k| e^{\hat{\Lambda}(t) }] \nonumber \\
 &=& \int \prod_k d^2\alpha_k 
 [ |a|^2 e^{-\frac{\beta\omega_0}{2}} \langle \alpha_k | e^{-\beta H_{+}} |\alpha_k \rangle +   
  |b|^2 e^{\frac{\beta\omega_0}{2}} \langle \alpha_k | e^{-\beta H_{-}} |\alpha_k \rangle ] 
  e^{ \sum_k [-2|g_k(t)|^2 -2g_k^{*}(t)\alpha_k^{*} + 2g_k(t)\alpha_k]}\nonumber \\
   \end{eqnarray}
Making use of standard Gaussian integration, we write 
\begin{eqnarray}
\label{A1}
&&\int \prod_k d^2\alpha_k \langle \alpha_k | e^{-\beta H_{+}} |\alpha_k \rangle
e^{ \sum_k [ -2g_k^{*}(t)\alpha_k^{*} + 2g_k(t)\alpha_k]}\nonumber \\
&=& \prod_k e^{\frac{\beta |g_k|^2}{\omega_k}} e^{-\frac{|g_k|^2}{\omega_k^2}(1-e^{-\beta \omega_k})}
\exp \left[ -\frac{[2g_k^{*}(t) + \frac{g_k^*}{\omega_k}(1-e^{-\beta \omega_k})]
[2g_k(t) - \frac{g_k}{\omega_k}(1-e^{-\beta \omega_k})] }{1-e^{-\beta \omega_k}}                    \right] \nonumber \\
\end{eqnarray}
and 
\begin{eqnarray}
\label{A2}
&&\int \prod_k d^2\alpha_k \langle \alpha_k | e^{-\beta H_{-}} |\alpha_k \rangle
e^{ \sum_k [ -2g_k^{*}(t)\alpha_k^{*} + 2g_k(t)\alpha_k]}\nonumber \\
&=& \prod_k e^{\frac{\beta |g_k|^2}{\omega_k}} e^{-\frac{|g_k|^2}{\omega_k^2}(1-e^{-\beta \omega_k})}
\exp  \left[ - \frac{[2g_k(t) + \frac{g_k}{\omega_k}(1-e^{-\beta \omega_k})]
[2g^*_k(t) - \frac{g^*_k}{\omega_k}(1-e^{-\beta \omega_k})] }{1-e^{-\beta \omega_k}}                     \right]\nonumber \\
\end{eqnarray}
Next we can calculate partition function and its value turns out to be
\begin{eqnarray}
 \label{Par}
 Z=\prod_k\Bigg(\frac{2\pi}{1-e^{-\beta\omega_k}}\Bigg ) \Bigg( |a|^2 e^{-\frac{\beta \omega_0}{2} }+ |b|^2 e^{\frac{\beta \omega_0}{2} } \Bigg)
\end{eqnarray}


Using equations (\ref{Par}),(\ref{A1}),(\ref{A2}),(\ref{fr}) in equation (\ref{rho1}), we write
\begin{eqnarray}
\label{coh}
 \rho_S(t) = \begin{pmatrix}
\frac{1+\langle\sigma_z\rangle}{2} & \langle\sigma_-\rangle  F(t)\\
\langle\sigma_+ \rangle F^*(t) & \frac{1-\langle\sigma_z\rangle}{2}
\end{pmatrix}
\end{eqnarray}
where 
\begin{eqnarray}
 F(t) &=&  \Bigg( \frac{|a|^2 e^{-\beta \omega_0/2} e^{ i \Phi(t)} +  |b|^2 e^{\beta \omega_0/2} e^{ -i \Phi(t)}  }
 {|a|^2 e^{-\beta \omega_0/2}+ |b|^2 e^{\beta \omega_0/2}} \Bigg)
  e^{-\gamma_1(t)}   
\end{eqnarray}
with  
\begin{eqnarray}
 \Phi(t)&=&\sum\limits_{k}\frac{4|g_k|^2}{\omega_k^2}\sin\omega_k t \\
 \gamma_1(t) &=&\sum\limits_{k} 4|g_k|^2\frac{1-\cos\omega_k t}{\omega_k^2}\coth\frac{\beta\omega_k}{2}
\end{eqnarray}
Since bath is of infinite degrees of freedom we could approximate the sums involved 
in the expression by integrals. Thus we have,
\begin{eqnarray}
\label{gt_nonmarkovian}
\Phi(t)&=&\int_0 ^\infty d\omega J(\omega)\frac{\sin(\omega t)}{\omega^2} \\
\gamma_1(t)&=&\int_0^\infty d\omega J(\omega)\frac{1-\cos\omega t}{\omega^2}\coth\frac{\beta\omega}{2}
\end{eqnarray}
Here, $J(\omega)$ is called spectral density function and it depends upon the kind of interaction between 
the bath and the system.

Using the fact $|a|^2+|b|^2=1$ and $\langle \sigma_z\rangle =|a|^2-|b|^2$, we can write 
\begin{eqnarray}
 \frac{|a|^2 e^{-\beta \omega_0/2} e^{ i \Phi(t)} +  |b|^2 e^{\beta \omega_0/2} e^{ -i \Phi(t)}  }
 {|a|^2 e^{-\beta \omega_0/2}+ |b|^2 e^{\beta \omega_0/2}}
 =\cos\Phi(t) - i \frac{ \sinh\frac{\beta \omega_0}{2} -\langle\sigma_z\rangle \cosh\frac{\beta \omega_0}{2}}
 {\cosh\frac{\beta \omega_0}{2}-\langle\sigma_z\rangle\sinh\frac{\beta \omega_0}{2}} \sin\Phi(t).
\end{eqnarray}
In order to get dephasing or time dependent frequency shift explicitly we define
\begin{equation}
\tan \chi(t) = \frac{\sinh(\beta \omega_0/2) -
\langle\sigma_z\rangle \cosh(\beta \omega_0/2)}{\cosh(\beta \omega_0/2)-
\langle\sigma_z\rangle \sinh(\beta \omega_0/2)}\tan \Phi (t) 
\end{equation}
so that $F(t)$ simplifies to $F(t) = e^{i\chi(t)} e^{-\gamma_1(t)-\gamma_2(t)}$ with 
\begin{eqnarray}
 \gamma_2(t) =\frac{1}{2} \ln \Bigg[ 1-\frac{(1-\langle\sigma_z\rangle^2)\sin^2 \Phi (t) }{\Big(\cosh(\beta \omega_0/2)
-\langle\sigma_z\rangle \sinh(\beta \omega_0/2)\Big)^2} \Bigg] 
\end{eqnarray}
with the help of these equations we can write $\sigma_+(t)$ defined in equation (\ref{coh}).

We note that for an initially uncorrelated state of the bath and system, we obtain \cite{Breuer}
\begin{eqnarray}
\label{DEN_final}
 \rho_S(t) = \begin{pmatrix}
\frac{1+\langle\sigma_z\rangle}{2}&\langle\sigma_-\rangle e^{-i\omega_0 t }e^{-\gamma_1(t)}\\
\langle\sigma_+\rangle e^{i\omega_0 t}e^{-\gamma_1(t)}&\frac{1-\langle\sigma_z\rangle}{2} 
\end{pmatrix} .
\end{eqnarray}

\subsection{Evaluation of terms in Master equation (\ref{SME})}
\label{appb}

In this section we provide derivation of the results presented above in Sec. \ref{variable}\ref{largespins}.
In order to simplify the calculations, we make the mapping from spin operators for spin-$\frac{1}{2}$  particles to hard-core bosons (HCBs). 
The HCBs are defined on lattice sites $i = 1, ..., N$ with restricted occupation numbers, $n_i= 0, 1$ \cite{Auerbach,Lone2016}. The constrained creation and destruction operators $b^{\dagger}_i$ and $b_i$,  are defined as  
$\sigma^+_i = b^{\dagger}_i$ , $\sigma^-_i = b_i$, and $\sigma_i^z=2b_i^{\dagger}b_i -1$. We then
observe that conservation of the total $ z $-spin $\sum_k \sigma^z_k $ implies conservation of total number of HCBs. 
The commutation properties of HCBs are given below: 
\begin{eqnarray}
[b_i,b_j]&=&[b_i,b^{\dagger}_j]= 0 , \textrm{ for } i \neq j , \nonumber\\
\{b_i,b^{\dagger}_i\}& = & 1 .
\label{commute}
\end{eqnarray}
Therefore, we can write the Hamiltonian in (\ref{hamspinbath}) as

\begin{align}
H_S & = \frac{\omega_r}{2} \sigma^z \nonumber \\
H_B & = \sum_k  \omega_k b^{\dagger}_k b_k \nonumber  \\
H_I & = \sum_k g_k [\sigma^+ b_k+ 
 \sigma^- b_k^{\dagger} + \sigma^z b_k^{\dagger} b_k]
 \label{hnew}
\end{align}
where $\omega_r = \omega_0 - \sum_k g_k  $ is the renormalized energy of system spin. 
We have dropped the $ 0 $ subscript on the system spin $ \sigma_0 \rightarrow \sigma $
as in this notation only this is denoted using Pauli matrices.  

Next we assume that the initial total density matrix has the form
$\rho(0) = \rho_S \otimes \rho_B $, and
write the non-Markovian quantum  master equation in the Schrodinger picture as \cite{Lone20162}
\begin{eqnarray}
 \frac{d \rho_S(t)}{dt}= -i[H_S,\rho_S]- \int_0^{t} \!\!\! d\tau 
{\rm Tr}_B [ H_I, [ 
H_I (t-\tau), \rho_S(t) \otimes \rho_B  ]    ].
\label{ME}
\end{eqnarray}
For a general polarized bath state we have
\begin{align}
\label{stab}
\rho_B & = | B(\theta) \rangle \langle B(\theta) |\nonumber \\
 | B(\theta) \rangle  & =\prod_{k=1}^N (c + d b_k^{\dagger})|0\rangle .
\end{align}
The first term in the double commutator can be written as
\begin{eqnarray}
\label{fterm}
 && \!\!\!\!\!\!\!\!\!\!\!\!\!\!\!\!\!\!\!\!
{\rm Tr}_{B}[H_I H_I(t-\tau)\rho_S\rho_B] \nonumber \\
 &=& \sum_k g_k^2 [\sigma^+\sigma^- 
 \rho_S \langle b_k b_k^{\dagger}\rangle_B e^{-i(\omega_0-\omega_k)(t-\tau)} + 
 \sigma^- \sigma^+ \rho_S \langle b^{\dagger}_k b_k \rangle_B e^{i(\omega_0 -\omega_k)(t-\tau)}   ] \nonumber\\
 &+& \sum_{kk^{\prime}} g_k g_{k^{\prime}}[ \langle b^{\dagger}_k b_k b^{\dagger}_{k^{\prime}} b_{k^{\prime}} \rangle_B \rho_S 
+ \sigma^+ \sigma^z \rho_S \langle b_k b^{\dagger}_{k^{\prime}} b_{k^{\prime}} \rangle_B
 + \sigma^- \sigma^z\rho_S \langle b^{\dagger}_k b^{\dagger}_{k^{\prime}} b_{k^{\prime}} \rangle_B  \nonumber \\
 &+& \sigma^z \sigma^+ \rho_S \langle b^{\dagger}_k b_k b_{k^{\prime}} \rangle_B e^{-i(\omega_0 -\omega_k)(t-\tau)}
 + \sigma^z \sigma^- \rho_S \langle b^{\dagger}_k b_k b^{\dagger}_{k^{\prime}} \rangle_B e^{i(\omega_0 - \omega_k)(t-\tau)} ]
 \nonumber \\
\end{eqnarray}
where $ \langle X \rangle_B = {\rm Tr} ( X \rho_B) $.
The various correlation functions above can be calculated using  Wick's theorem. However, if we make transformation as defined in 
(\ref{basistrans}), such that all sites are filled ($ c = 0, d= 1 $), then the correlation functions with odd number of HCB operators vanish. In this case we can  write (\ref{fterm}) as
\begin{eqnarray}
 \int_0^{t} d\tau {\rm Tr}_B [H_I H_I(t-\tau)\rho_S\rho_B] = 
 \Gamma_1(t) \sigma^-\sigma^+ \rho_S +\Gamma_d(t) \rho_S  
 \end{eqnarray}
where 
\begin{eqnarray}
\Gamma_d(t) &=& (\sum_k g_k)^2t \\
 \Gamma_1(t) &=&  \sum_k g_k^2 \int_0^t dt e^{i(\omega_0-\omega_k)(t-\tau)} 
\end{eqnarray}
Similarly,
\begin{eqnarray}
 \int_0^{t} d\tau {\rm Tr}_B [H_I \rho_S\rho_B H_I(t-\tau)] &=& \Gamma_1^{*}(t) \sigma^+ \rho_S \sigma^- 
   + \Gamma_d(t)  \sigma^z \rho_S \sigma^z \nonumber \\
 \int_0^{t} d\tau {\rm Tr}_B [  H_I(t-\tau) \rho_S\rho_B  H_I] &=&  
 \Gamma_1(t) \sigma^+ \rho_S \sigma^- 
 + \Gamma_d(t)  \sigma^z \rho_S \sigma^z \nonumber \\
  \int_0^{t} d\tau {\rm Tr}_B [\rho_S\rho_BH_I H_I(t-\tau)] &=& 
  \Gamma_1^*(t) \rho_S \sigma^-\sigma^+  +\Gamma_d(t) \rho_S \nonumber .
\end{eqnarray}
Using all the above equations  in the master equation (\ref{ME}), we obtain 

\begin{eqnarray}
 \frac{d \rho_S }{dt} &=& -i [\tilde{H}(t),\rho_S]
 + 2\Gamma_{d}(t)[\sigma^z\rho_S\sigma^z-\rho_S] \nonumber\\
 && + \Gamma_0(t)[2\sigma^+ \rho_S \sigma^- -\sigma^+ \sigma^- \rho_S - \rho_S\sigma^+ \sigma^-] ,
 \end{eqnarray}
where
\begin{eqnarray}
\tilde{H}(t) &=& \frac{\omega_r}{2}\sigma^z + \sum_k g_k^2 \frac{1-\cos(\omega_0-\omega_k)t}{\omega_0-\omega_k} \sigma^-\sigma^+ \\
 \Gamma_0(t) &=&  \sum_k g_k^2 \frac{\sin(\omega_0-\omega_k)t}{\omega_0-\omega_k} . 
\end{eqnarray}

\twocolumngrid



\end{document}